\documentclass[%
 reprint,
 amsmath,amssymb,
 aps,
 prd,
]{revtex4-1}

\usepackage{graphicx}
\usepackage{dcolumn}
\usepackage{bm}

\def\riNa{^{22}{\rm Na}}
\def\riCf{^{252}{\rm Cf}}

\def\Leff{\mathscr{L}_{\rm eff}}
\def\s2s1{\log_{10}({\rm S2/S1})}
\usepackage{mathrsfs}
\usepackage{multirow}

\begin{document}

\preprint{APS/123-QED}

\title{Measurement of the scintillation efficiency for nuclear recoils in liquid argon\\ under electric fields up to $3\ {\rm kV/cm}$}

\author{M.Kimura}
 \email{masato@kylab.sci.waseda.ac.jp}

\author{M.Tanaka}

\author{T.Washimi}
 \altaffiliation[Present address: ]{National Astronomical Observatory of Japan, Mitaka, Tokyo, Japan}

\author{K.Yorita}
 \email{kohei.yorita@waseda.jp}
\affiliation{%
Waseda University, 3-4-1, Okubo, Shinjuku, Tokyo, 169-8555, Japan
}%


\date{\today}

\begin{abstract}
We present a measurement of scintillation efficiency for a few tens of keV nuclear recoils (NR) with a liquid argon time projection chamber under electric fields ranging from $0$ to $3\ {\rm kV/cm}$.
The calibration data are taken with $\riCf$ radioactive source.
Observed scintillation and electroluminescence spectra are simultaneously fit with spectra derived from Geant4-based Monte Carlo simulation and an NR model.
The scintillation efficiency extracted from the fit is reported as a function of recoil energy and electric field.
This result can be used for designing the detector and for the interpretation of experimental data in searching for scintillation and ionization signals induced by WIMP dark matter.
\end{abstract}

\maketitle

\section{\label{sec:Intro}Introduction}
Dark matter appears to be spread across galaxies through astronomical and cosmological observations, and many groups have been trying to detect it directly using a variety of detector techniques and target materials.
There are many direct detection experiments to identify nuclear recoils (NR) induced by the elastic scattering of weakly interacting massive particle (WIMP) off target nuclei.
The magnitude of the typical recoil energy in these experiments is a few tens of keV.
A liquid argon (LAr) time projection chamber (TPC) is known to offer several attractive features to search for a WIMP with a mass in the GeV to TeV  range \cite{benetti2008first, amaudruz2018first, agnes2018darkside, agnes2018Low, ajaj2019search}:
the efficient conversion of energy deposition into scintillation and ionization signals, a powerful discrimination of NR signal from an electronic recoils (ER) background, and a reasonably high recoil energy for WIMP-Ar nuclear scattering due to relatively small atomic mass of argon.
In the double-phase (liquid/gas) TPC, excitation and ionization are induced by an incident particle after interacting with LAr, leading to a prompt scintillation signal (S1).
The active volume of the detector is subjected to a uniform electric field, which causes the ionization electrons to escape recombination and drift toward the gaseous region, where they emit an electroluminescence signal (S2).
It is known that light and charge yields, i.e., the number of scintillation photons and the number of ionization electrons, respectively, produced by an NR of a given energy, depend on both the energy and applied electric field.
The scintillation efficiency $\Leff$ is defined as the light yield for NR per recoil energy relative to that of ER measured at a null field.
Although it has been measured by several groups \cite{brunetti2005warp, gastler2012measurement, cao2015measurement, creus2015scintillation, agnes2017simulation, agnes2018measurement}, the properties for electric fields greater than $1\ {\rm kV/cm}$ have not been explicitly discussed yet.
In this work, we report the first measurement of the scintillation efficiency $\Leff$ resulting from a few tens of keV of NR under electric fields up to $3\ {\rm kV/cm}$.

\section{\label{sec:NRModel}Nuclear recoils model}
\subsection{\label{subsec:NRModel_Framework}Framework}
\begin{figure}[tb]
	\centering
	\includegraphics[width=1.0\columnwidth]{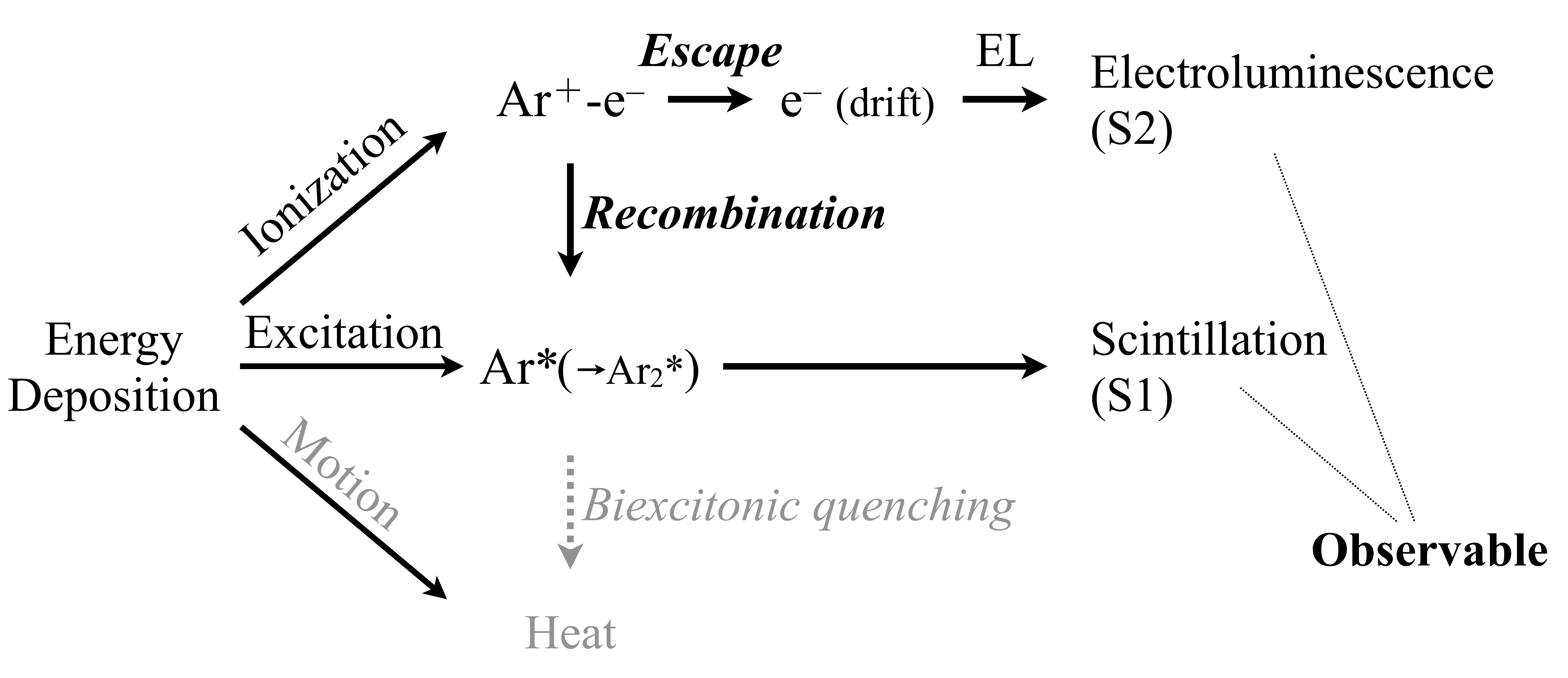}
	\caption{Schematic for the conversion process of energy deposition into observables.
	Energy deposition is distributed to three channels: ionization, excitation, and atomic motion.
	The excitations lead to S1, the ionization electrons lead to S2, and the atomic motion is unobservable in LAr-TPC.
	Through the recombination process, a ratio of S1 and S2 changes.}
	\label{fig:NRModel}
\end{figure}

The observable quantities in the double-phase LAr-TPC are S1 and S2 light signals.
A schematic for the conversion process of energy deposition into these observable quantities is shown in Fig. \ref{fig:NRModel}.
An energy deposition of $E_0$ is distributed as expressed in the following equation:
\begin{equation}
	\label{eq:EnergyDestribution}
	\frac{E_0 L}{W} = N_{ex} + N_i = N_i (\alpha +1)
\end{equation}
where $W = 19.5\ {\rm eV}$ is the effective work function \cite{doke1988let}, $L$ is an additional factor for NR that accounts for energy loss due to atomic motion,  $N_{ex}$ and $N_i$ are the average number of produced excitons and electron-ion pairs, respectively, and $\alpha$ is the exciton-to-ion ratio.
The factor $L$ is predicted using the Lindhard theory \cite{lindhard1963integral} as follows:
\begin{eqnarray}
	\label{eq:Lindhard}
	L &=& \frac{kg(\epsilon)}{1+kg(\epsilon)}, \\ 
	k &=& 0.133Z^{2/3}A^{-1.2}, \nonumber \\
	g(\epsilon) &=& 3\epsilon^{0.15} + 0.7\epsilon^{0.6}+\epsilon , \nonumber \\
	\epsilon &=& 11.5 (E_{0}/ {\rm keV}) Z^{-7/3}, \nonumber
\end{eqnarray}
where $Z = 18$ and $A = 40$ are the atomic and mass numbers, respectively.
The ratio $\alpha$ is parametrized as an empirical function of the electric field $F$, similar to the description for liquid xenon in Ref. \cite{lenardo2015global},
\begin{equation}
	\label{eq:alpha_def}
	\alpha = \alpha_0 \exp(-D_\alpha F),
\end{equation}
where $\alpha_0$ and $D_\alpha$ are free parameters.
Once the energy deposition is distributed to ionization, excitation, or atomic motion channels, all the excitons contribute directly to the emission of scintillation photons.
A fraction of electrons recombines with ions to produce additional scintillation photons, whereas the rest of the electrons become ionization electrons.
The electrons drift toward the gaseous region and emit electroluminescence.
An empirical modification \cite{joshi2014first} of the Thomas-Imel box model (TIB model) \cite{thomas1987recombination} provides the recombination probability $R$ as follows:
\begin{eqnarray}
	\label{eq:TIBmodel}
	R &=& 1-\frac{\ln(1+N_i \varsigma)}{N_i  \varsigma}, \\
	\varsigma &=& \gamma F^{-\delta}. \nonumber \nonumber
\end{eqnarray}
Here, $\gamma$ and $\delta$ are free parameters.
Biexcitonic quenching, where a collision of two excitons produces a single photon, is incorporated by the Mei model \cite{mei2008model}.
The quenching term $f_l$ is parametrized by
\begin{equation}
	\label{eq:BirksLaw}
	f_l = \frac{1}{1+k_B (\frac{dE}{dx})_{\rm el}},
\end{equation}
where $k_B$ is a free parameter. 
The electronic stopping power $(\frac{dE}{dx})_{\rm el}$ is given by Mei {\it et al.} as a function of the recoil energy $E_0$ (Fig. 5 in Ref. \cite{mei2008model}).
Summarizing these effects, the number of produced scintillation photons, $n_{ph}$, and the number of produced ionization electrons, $n_e$, are expressed as follows:
\begin{eqnarray}
	\label{eq:nph_as_Enr}
	n_{ph} &=& L \times f_l \times \frac{E_{0}}{W} \times \biggl[ 1- \Bigl( \frac{1}{1+\alpha} \Bigr) \ (1-R) \biggr], \\
	\label{eq:ne_as_Enr}
	n_e &=& L \times \frac{E_{0}}{W} \times \Bigl( \frac{1}{1+\alpha} \Bigr) \ (1-R) .
\end{eqnarray}
These quantities are related to S1 and S2 as follows:
\begin{eqnarray}
	\label{eq:S1_as_nph}
	{\rm S1} &=& g_1 n_{ph}, \\
	\label{eq:S2_as_ne}
	{\rm S2} &=& g_2 n_e,
\end{eqnarray}
where $g_1$ is the scintillation photon collection efficiency and $g_2$ is the average number of detected electroluminescence photons per one drift electron.
Both $g_1$ and $g_2$ are considered as detector properties and remain constant for NR and ER events.

\subsection{\label{subsec:NRModel_Reconstruction}Fitting procedure}
For NR at a null field, $R$ is expected to be $1$; therefore, $k_B$ is the only free parameter to account for the quenching.
As applying the electric field, $R$ is expected to decrease, resulting in the suppression of S1 signal and production of more S2 signal.
The related parameters of this process are $\alpha_0 , D_\alpha , \gamma ,$ and $\delta$ in Eqs. (\ref{eq:alpha_def}) and (\ref{eq:TIBmodel}).
In the previous measurements of the light yields, a value of $\alpha \sim 1$ is suggested to describe the observed data \cite{cao2015measurement, agnes2018measurement}.
We interpret the value to the approximation at the lower electric field and constrain $\alpha_0$ to $1$.
We first determine $k_B$ from the S1 spectrum of the null field data sample, and then $D_\alpha , \gamma ,$ and $\delta$ are obtained from both S1 and S2 spectra under electric fields.

In this measurement, $511\ {\rm keV}$ $\gamma$ ray line from a $\riNa$ source is used as the reference ER events of the scintillation efficiency $\Leff$.
By definition of the effective work function $W$, the observed S1 light signal (${\rm S1_{\rm Na}}$) by the energy deposition of $E_{\rm Na} = 511\ {\rm keV}$ from ER at null field is represented as following:
\begin{equation}
	\label{eq:S1_as_Ee}
	{\rm S1_{\rm Na}} = g_1 \frac{E_{\rm Na}}{W}.
\end{equation}
Equation (\ref{eq:S1_as_Ee}) implies $L=f_l=R=1$ for ER at a null field, which comes from the definition of the $W$ rather than the true underlying values of these effects.
From Eqs. (\ref{eq:nph_as_Enr}), (\ref{eq:S1_as_nph}), and (\ref{eq:S1_as_Ee}), the scintillation efficiency $\Leff$ referenced to the $511\ {\rm keV}$ $\gamma$-ray line of a $\riNa$ source is given by
\begin{equation}
	\label{eq:Leff_def}
	\Leff (E_0, F) = \frac{({\rm S1}/E_0)}{({\rm S1}_{\rm Na}/E_{\rm Na})} = \frac{n_{ph}}{E_0/W}.
\end{equation}

A value of $g_1 = 0.12 \pm 0.01$ is measured in this work and will be described in Sec. \ref{subsec:Apparatus_calibration}.
We derived $g_2/g_1$ value by analyzing the ER data samples under electric fields that were taken by the same experimental condition \cite{washimi2018phd}.
A value of $g_2/g_1 = 10 \pm 2$ is obtained and used as the detector constants.

\section{\label{sec:Apparatus}Apparatus}
\subsection{\label{subsec:Apparatus_setup}Detector and geometry}
\begin{figure}[t]
	\centering
	\includegraphics[width=1.0\columnwidth]{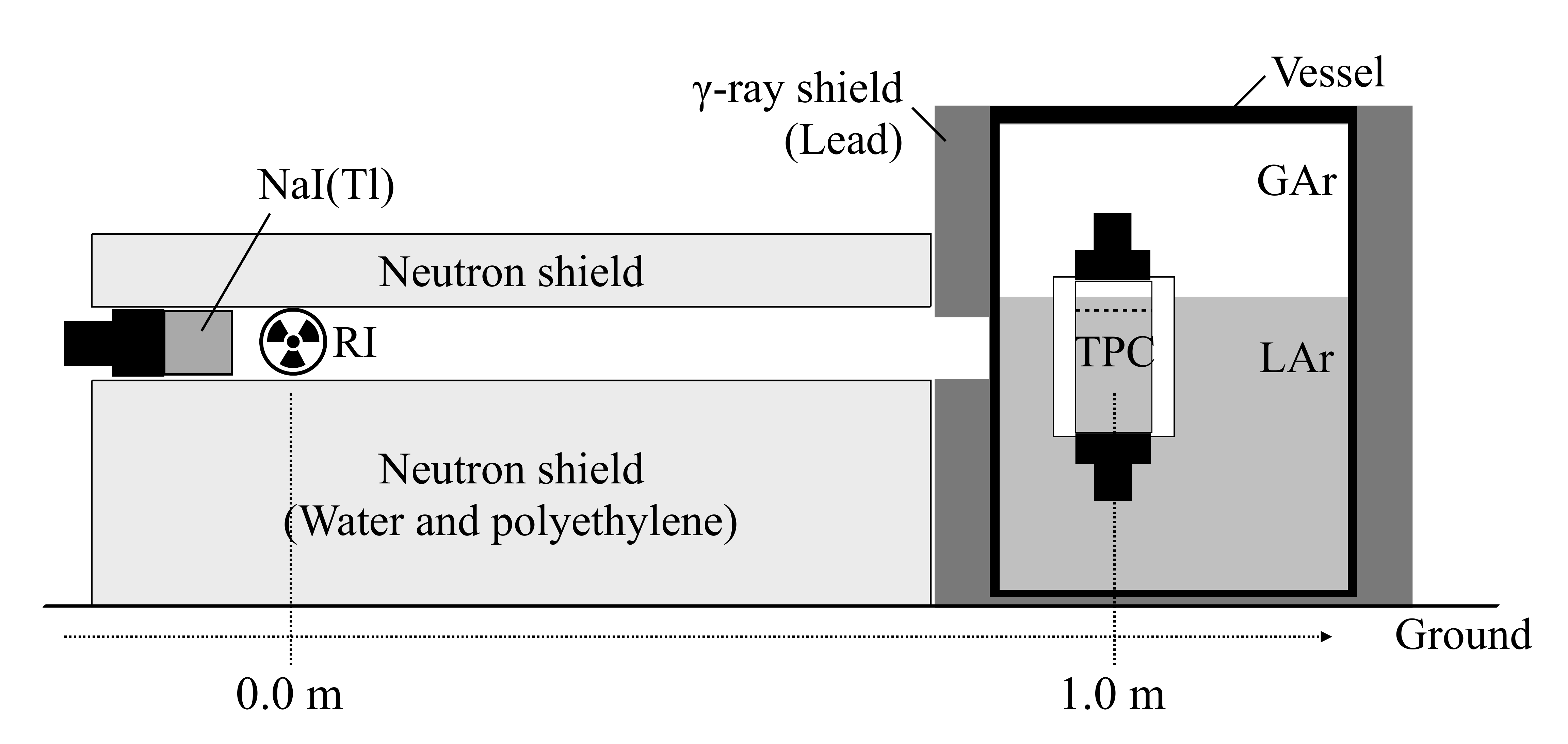}
	\caption{Schematic of the experimental apparatus.}
	\label{fig:Apparatus}
\end{figure}
We measured the scintillation efficiency at the LAr test stand at Waseda University \cite{tanaka2013status}.
The double-phase TPC used in this study has an active region of diameter of $6.4\ {\rm cm}$ and height of $10\ {\rm cm}$ with two PMTs (photomultiplier tubes)  (HAMAMATSU R11065).
An extraction grid is placed at the top of the active region.
The gap between the extraction grid and anode is $1\ {\rm cm}$, and the liquid surface is kept centered between them.
A Cockcroft-Walton circuit is mounted in the LAr that surrounds the TPC to supply high voltage to the electrodes of TPC.
Data were taken under electric fields of $0.0, 0.2, 0.5, 1.0, 2.0,$ and $3.0\ {\rm kV/cm}$.
More details are described elsewhere \cite{tanaka2013status, washimi2018scintillation, kimura2019status}.

Figure \ref{fig:Apparatus} shows a schematic of the experimental apparatus used in this measurement.
A $\riCf$ neutron source with a spontaneous fission rate of approximately $1 \times 10^5\ {\rm fission/s}$ is placed at a distance of $1.01 \pm 0.01\ {\rm m}$ from the center of the TPC.
A NaI(Tl) scintillator ($2\ {\rm in.} \times \ 2\ {\rm in.}$ cylinder) located beside the neutron source provides timing information by detecting associated $\gamma$ ray.
A lead shield surrounds the vessel to suppress background from ambient $\gamma$ rays.
Other background arises because of neutrons from the $\riCf$ source; this background from the $\riCf$ source reaches the active region via a single or multiple scattering at any part of the materials in the laboratory.
Water and polyethylene shields are placed to suppress these scattered neutrons.
The data acquisition is triggered by the coincidence between the TPC PMTs and NaI(Tl) scintillator signals within a $1\ {\rm \mu s}$ window.
Both TPC PMT and NaI(Tl) scintillator waveforms are digitized at a frequency of $250\ {\rm MHz}$ using a flash ADC (SIS3316).
The length of the digitizer records is set to a value long enough to detect S2 of the maximum drift ($10\ {\rm cm}$) events.

\subsection{\label{subsec:Apparatus_calibration}Energy calibration of TPC}
\begin{figure}[tb]
	\centering
	\includegraphics[width=1.0\columnwidth]{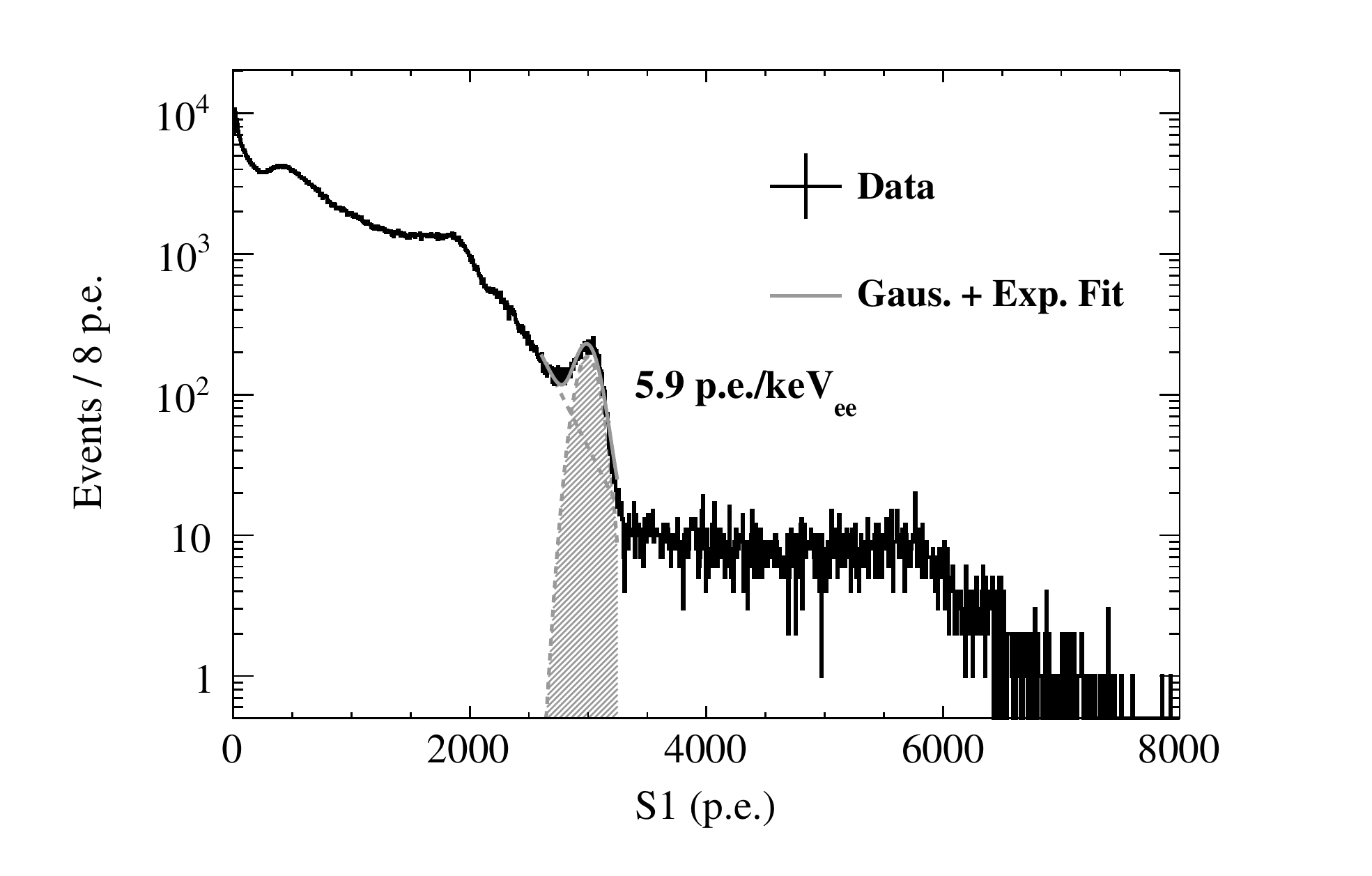}
	\caption{$\riNa$ spectrum for the energy calibration at a null field, with the Gaussian plus exponential fitting around the $511\ {\rm keV}$ full absorption peak.}
	\label{fig:22NaLY}
\end{figure}
Figure \ref{fig:22NaLY} shows the S1 spectrum of the $\riNa$ data taken at a null field.
We determine the observed S1 signal per ER energy at a null field, ${\rm S1}_{\rm Na}/E_{\rm Na}$, as $5.9 \pm 0.3\ {\rm p.e./keV_{ee}}$ (p.e.: photoelectron, ee: electron equivalent) by fitting $511\ {\rm keV}$ full absorption peak with a Gaussian plus exponential function.
The corresponding scintillation photon collection efficiency $g_1$ of the detector is $0.12 \pm 0.01$.

\section{Event reconstruction}
\label{sec:EventReco}
\begin{figure}[tb]
	\centering
	\includegraphics[width=1.0\columnwidth]{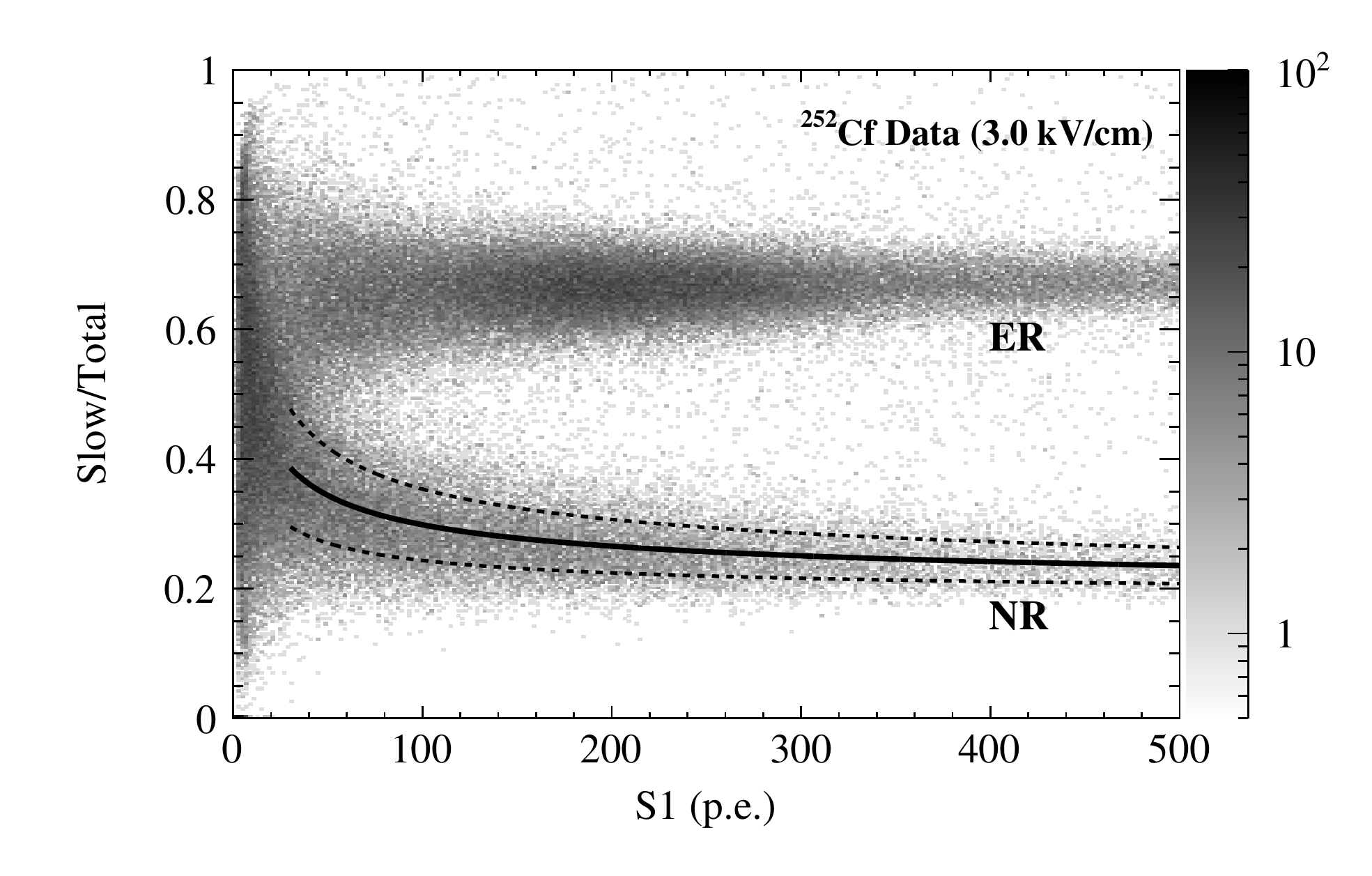}
	\caption{Distribution of the PSD parameter (${\rm slow/total}$) versus S1 with data taken with the $\riCf$ neutron source under the electric field of $3\ {\rm kV/cm}$.
	Two dashed lines correspond to the band for PSD cut ($\pm 1\sigma$).}
	\label{fig:S1vsPSD}
\end{figure}
The energy of the incident neutron from the $\riCf$ source is reconstructed based on time of flight (TOF), i.e., the time difference between the NaI(Tl) and TPC signals.
The arrival time of a pulse is identified as the first digitized sample above a threshold of $50\%$ peak amplitude.
The S1 is reconstructed as an integrated charge in the time interval between $-0.04$ and $5.0\ {\rm \mu s}$.
The pulse shape discrimination (PSD) parameter slow/total is defined as the fraction of light detected after $0.12\ {\rm \mu s}$ of the S1 signal.
The S2 is reconstructed as an integrated charge after $10\ {\rm \mu s}$ in the data acquisition window.
Events of the data samples are selected by requiring one proper S1 pulse.
For data samples taken under the electric fields, the additional requirement to have one proper S2 pulse is applied to select single scattered NR events.
Figure \ref{fig:S1vsPSD} shows a distribution of the PSD parameter (slow/total) versus S1 with data taken under the electric field of $3\ {\rm kV/cm}$, after requiring the TOF to be in the range of $43$-$111\ {\rm ns}$, corresponding to an incident neutron energy of $0.41$-$2.44\ {\rm MeV}$.
The neutrons from the $\riCf$ source can induce a $\gamma$ ray through an interaction with passive detector materials.
These induced $\gamma$ rays are observed to result in ER events having neutronlike timing.
Thus, a PSD band cut ($\pm 1\sigma$) is imposed to select NR events and suppress ER contamination.
A contribution from accidental coincidence backgrounds is estimated from a negative TOF window of $-0.9$ to $-0.2\ {\rm \mu s}$.

\section{\label{sec:Method}Method}
\subsection{\label{subsec:Method_MC}Monte Carlo}
\begin{figure}[t]
	\centering
	\includegraphics[width=1.0\columnwidth]{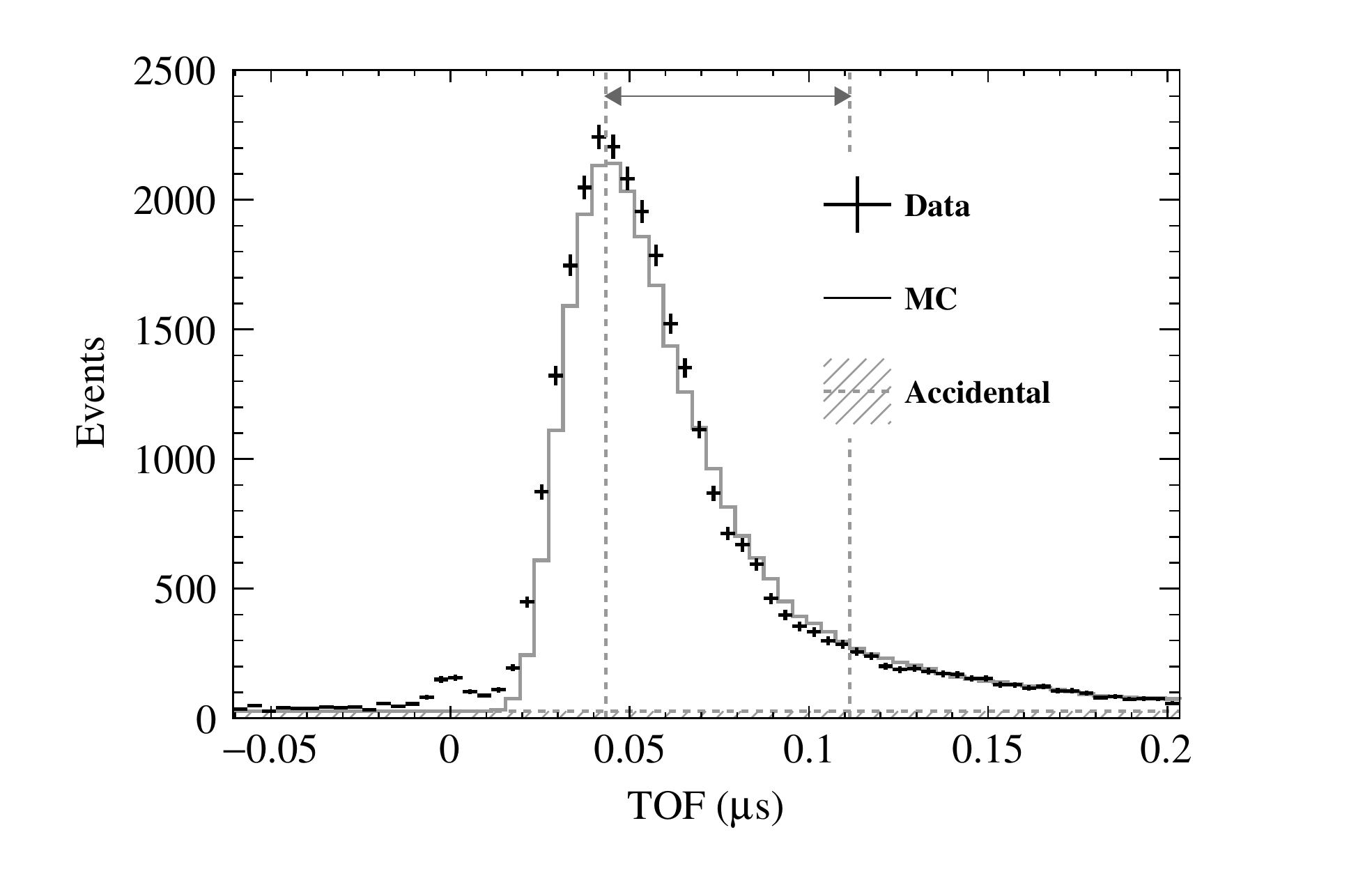}
	\caption{Comparison between data and the MC TOF spectra.
	The vertical dashed lines and gray arrow represent the TOF range where the simultaneous fit is performed.
	Contamination from ER events that are not simulated in MC produces the peak around $0\ {\rm s}$.}
	\label{fig:TOFwAc_nullE}
\end{figure}
\begin{figure}[t]
	\centering
	\includegraphics[width=1.0\columnwidth]{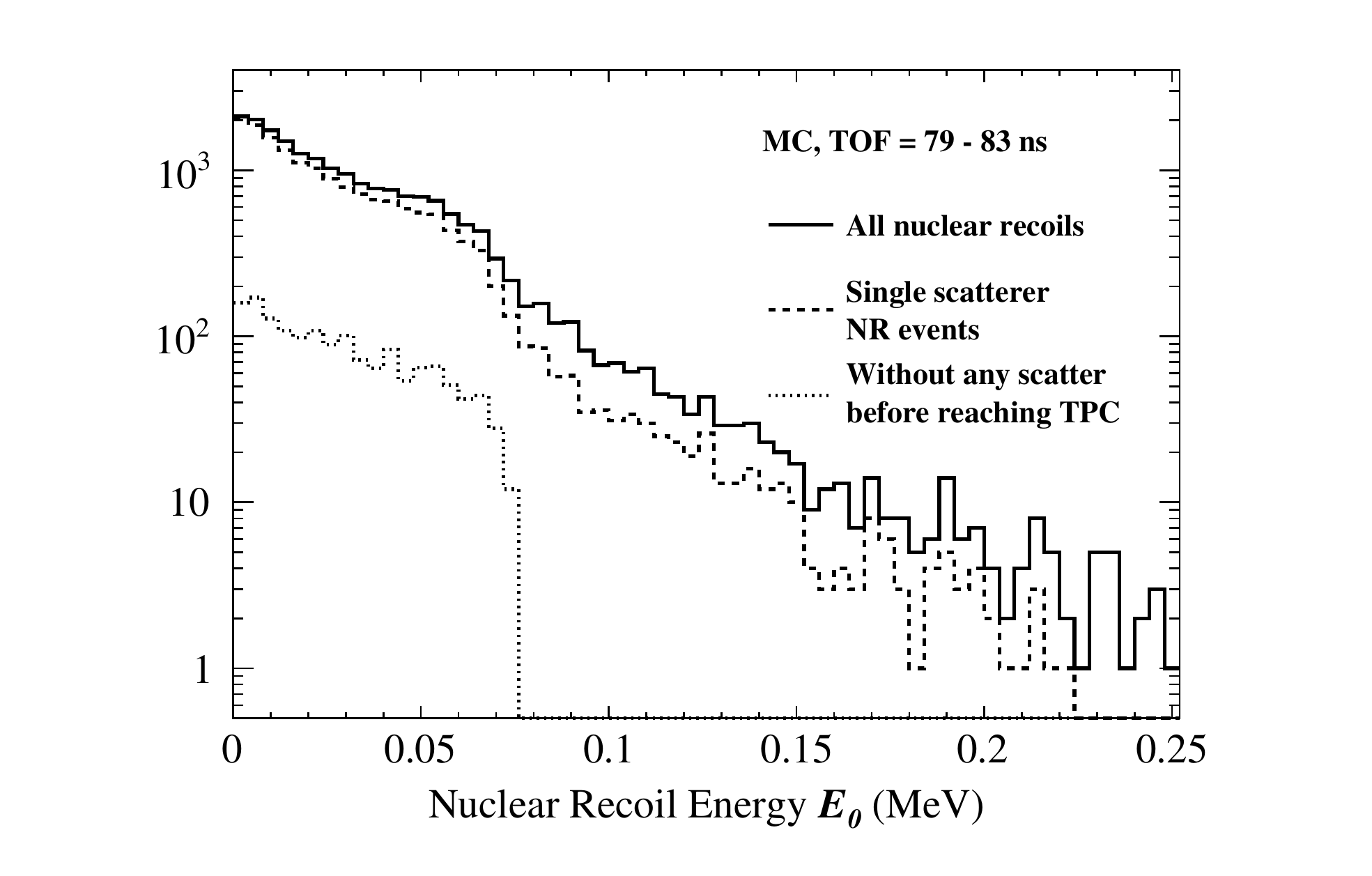}
	\caption{Energy deposition spectra derived from Geant4-based MC simulation.
	Shown are all NR in the LAr active region (solid line), contributions from single scattered NR events (i.e., neutrons that scattered only once in the active region) (dashed line), and neutrons that reached without any scattering in any part of the apparatus before reaching the active region (dotted line).}
	\label{fig:MCNRProperty}
\end{figure}

Energy deposits by the neutrons are simulated in a Geant4-based \cite{agostinelli2003geant4, allison2006geant4} Monte Carlo (MC) simulation of the experimental apparatus, using a neutron spectrum of $\riCf$ in Ref. \cite{boldeman1986measurements} and nuclear data library files G4NDL 4.5 \cite{mendoza2014new, mendoza2012new, mendoza2018update} with revised differential cross sections for elastic scattering from Ref. \cite{robinson2014new}.
We confirmed the validity of the simulation using a comparison of the TOF distribution between data and MC in Fig. \ref{fig:TOFwAc_nullE}.
The observed events in data at around ${\rm TOF} = 0\ {\rm s}$ mainly consist of low-energy (${\rm S1} \lesssim 30\ {\rm p.e.}$) events.
These events are considered as contamination from ER events and not used in this analysis.
Figure \ref{fig:MCNRProperty} shows the energy deposition ($E_0$) distribution from the MC simulation for the TOF range of $79$-$83\ {\rm ns}$, corresponding to a neutron energy of about $0.75\ {\rm MeV}$.
While the $\riCf$ source has a continuous neutron spectrum, a backscatter edge would be visible by constraining the TOF.
The edge of each TOF bin is useful to resolve degeneracy between the free parameters as described later.
The leading contribution is expected from the neutrons that are scattered more than once in any part of the apparatus (such as neutron/gamma shieldings, the vessel, and the LAr that surrounds the TPC) before reaching the active region.
However, the position of the backscatter edge is not affected, as shown in Fig. \ref{fig:MCNRProperty}.

\subsection{\label{subsec:Method_Fit}Data fitting}
\begin{figure}[t]
	\centering
	\includegraphics[width=1.0\columnwidth]{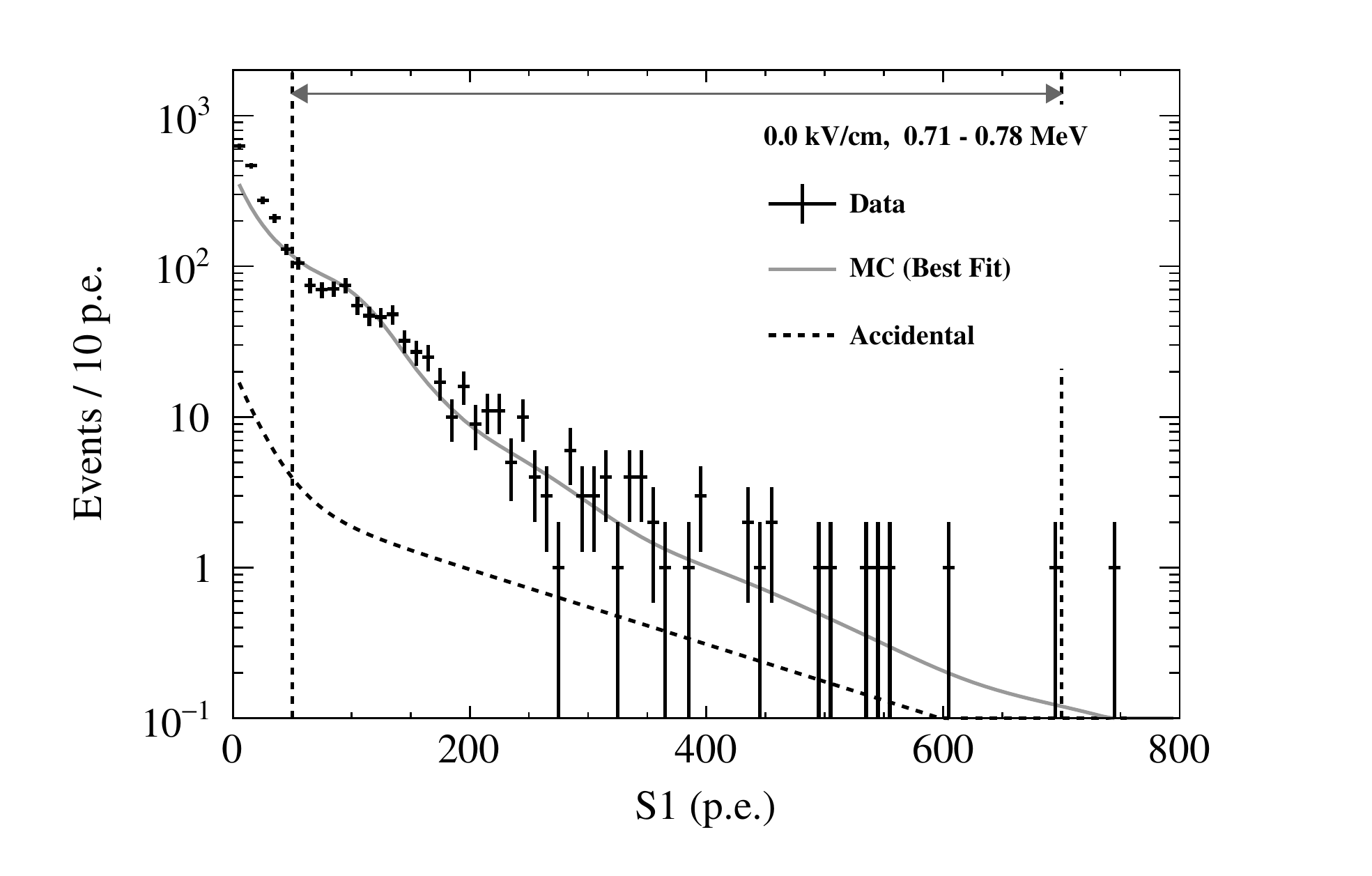}
	\caption{S1 spectrum of the NR data sample taken at a null field and MC-derived spectrum simultaneously fitted to experimental data for TOF in the range of $79$-$83\ {\rm ns}$.
	The area indicated by vertical dashed lines and gray arrow represents the fitting range.}
	\label{fig:S1FitnullE_79_83ns}
\end{figure}
\begin{figure}[t]
	\includegraphics[width=1.0\columnwidth]{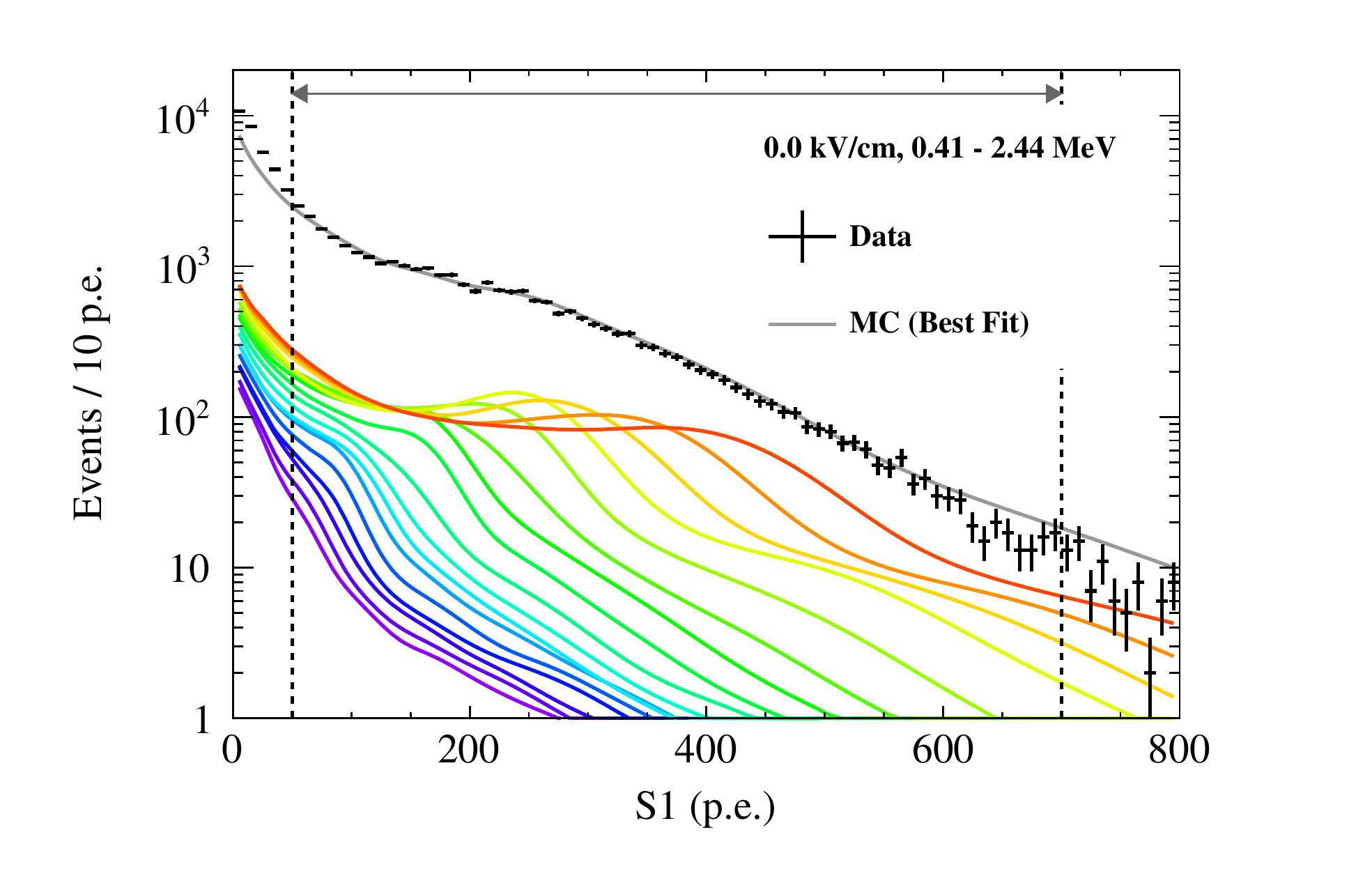}
	\caption{S1 spectrum of the NR data sample taken at a null field (black point) and MC-derived spectrum simultaneously fitted to experimental data (gray line) for the entire TOF range of interest.
	Also shown are MC-derived spectra (colored lines) representing the contribution of each TOF bin, from $43$-$47\ {\rm ns}$ (red) to $107$-$111\ {\rm ns}$ (violet).
	The area indicated by vertical dashed lines and gray arrow represents the fitting range.}
	\label{fig:S1FitnullE_AllBin}	
\end{figure}
\begin{figure}[!tb]
	\centering
	\includegraphics[width=1.0\columnwidth]{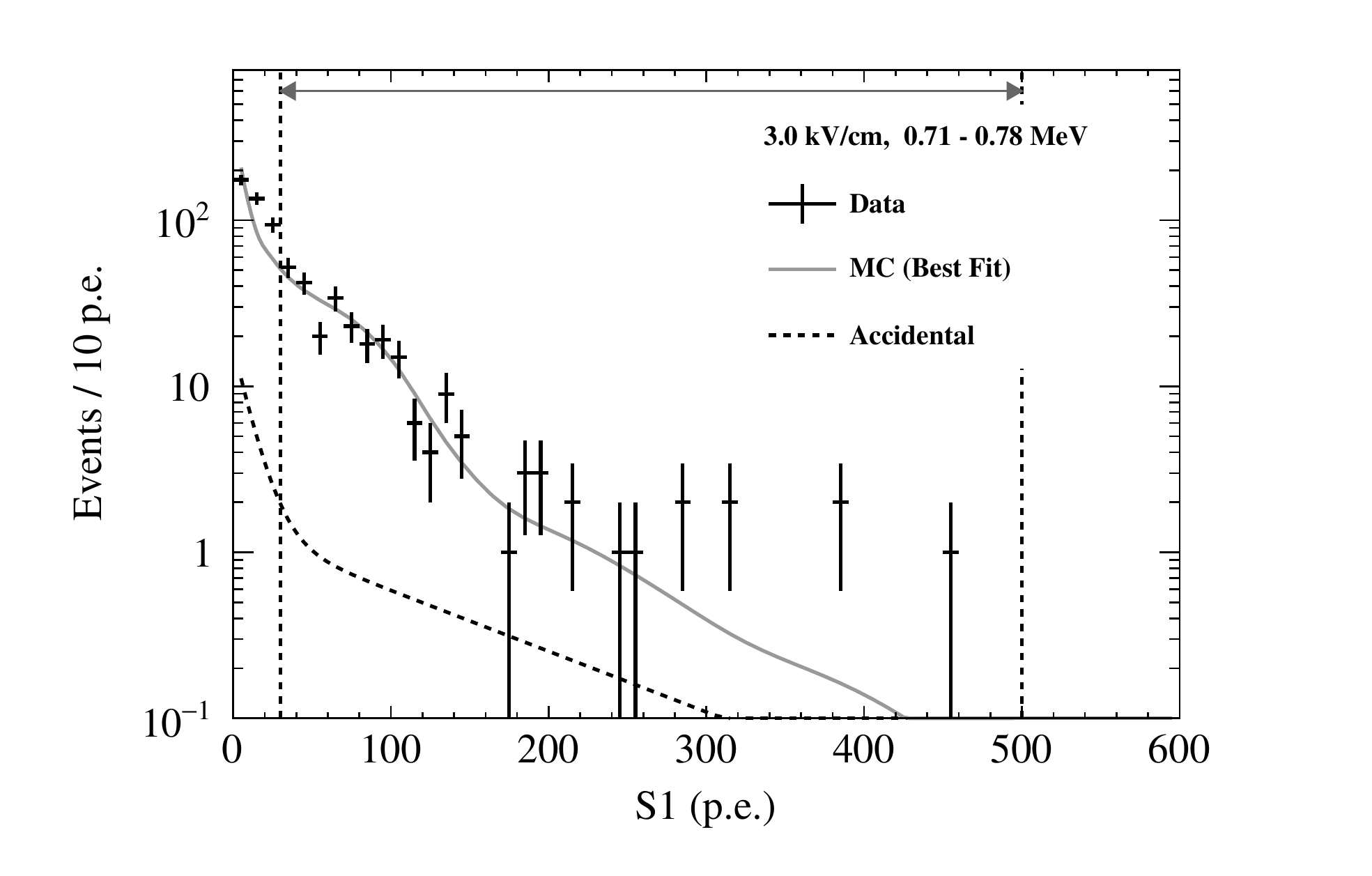}
	\includegraphics[width=1.0\columnwidth]{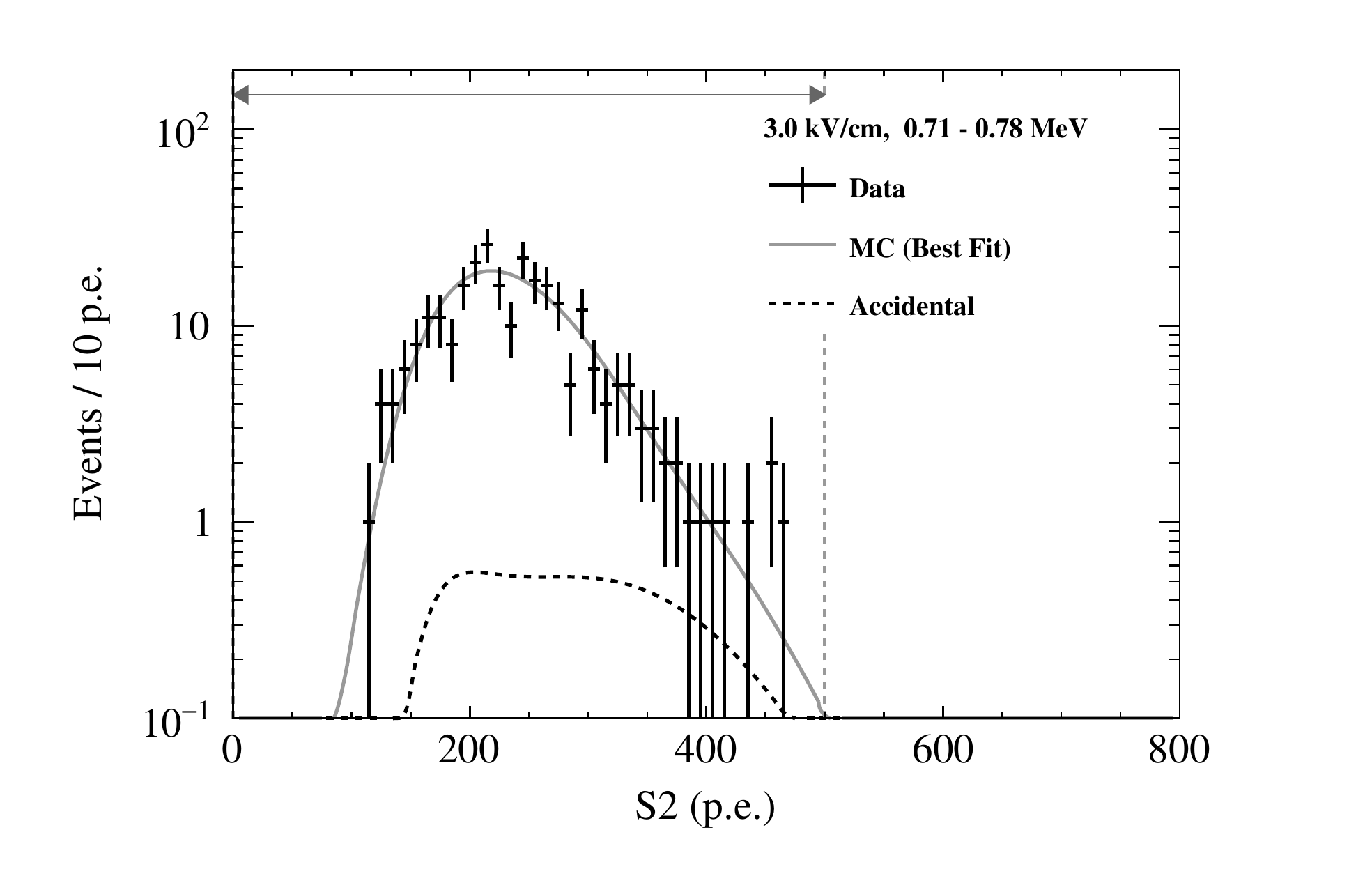}	
	\caption{S1 (top) and S2 (bottom) spectra of the NR data sample taken at the electric field of $3.0\ {\rm kV/cm}$ and MC-derived spectra simultaneously fitted to experimental data for TOF in the range of $79$-$83\ {\rm ns}$.
	The areas indicated by vertical dashed lines and gray arrow represent the fitting range.}
	\label{fig:S1andS2Fit3kVcm_79_83ns}
\end{figure}
\begin{figure}[!tb]
	\centering
	\includegraphics[width=1.0\columnwidth]{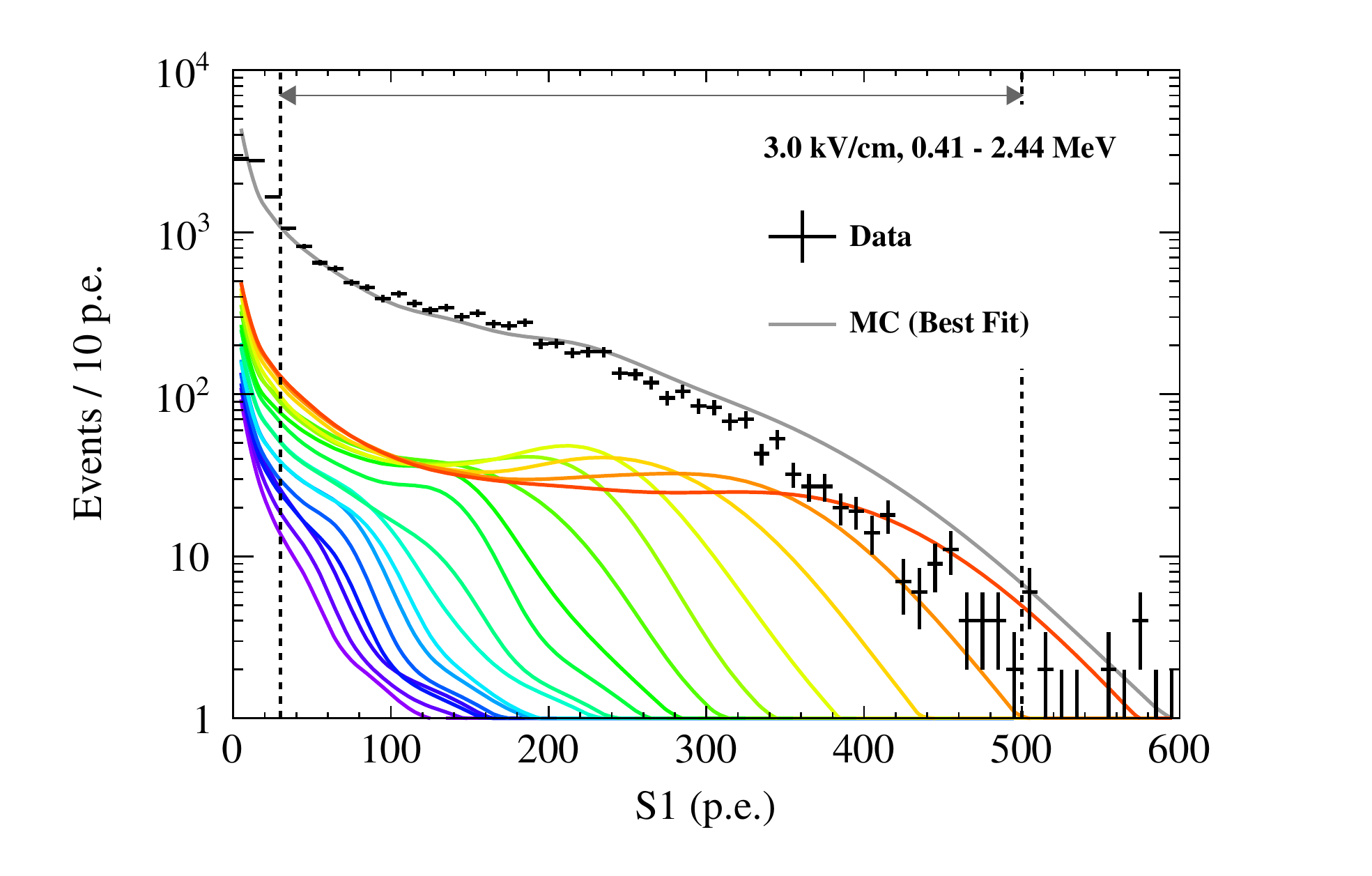}
	\includegraphics[width=1.0\columnwidth]{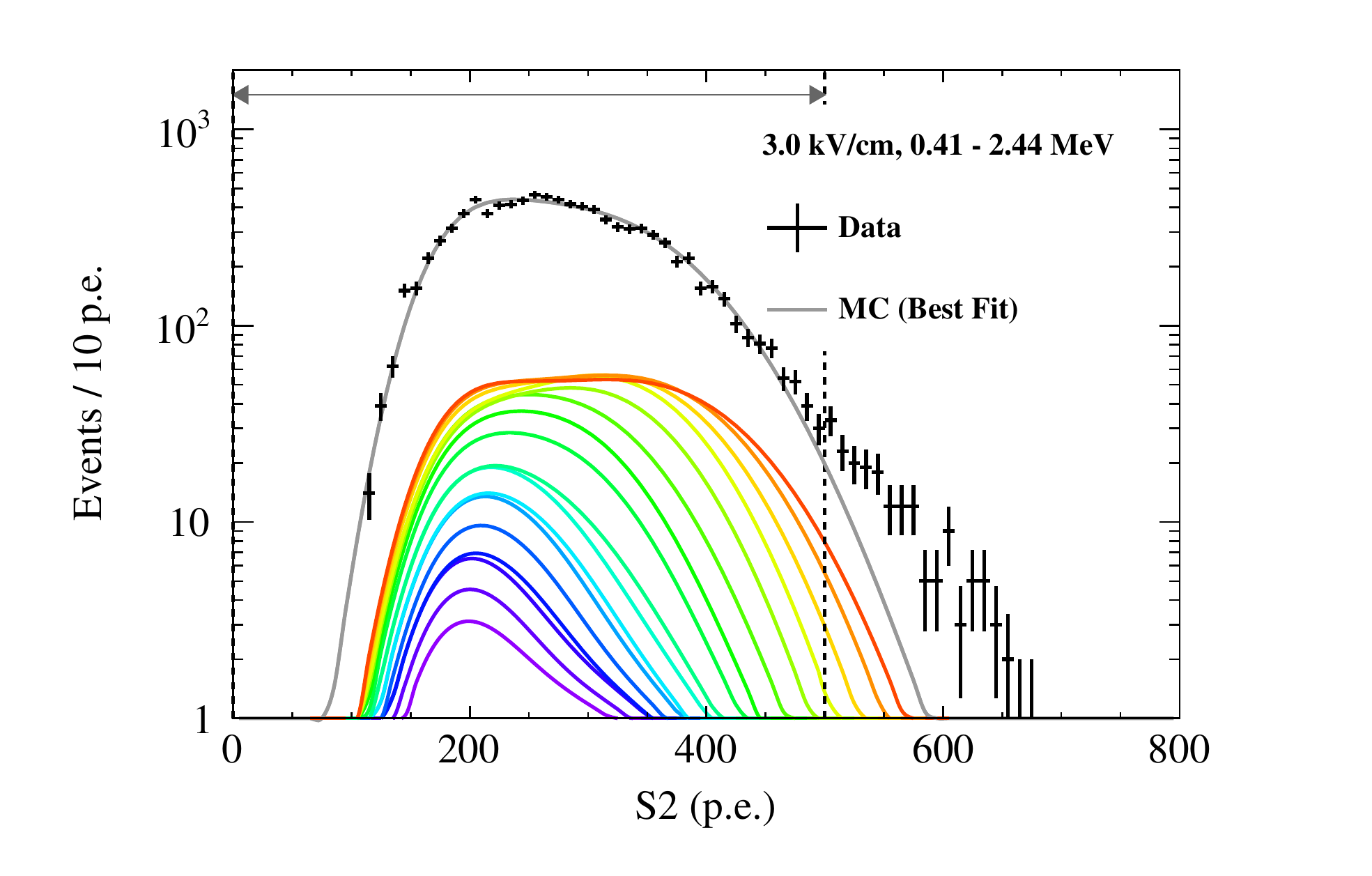}	
	\caption{S1 (top) and S2 (bottom) spectra of the NR data sample taken at the electric field of $3.0\ {\rm kV/cm}$ and MC-derived spectra simultaneously fitted to experimental data for the entire TOF range of interest.
	The figure description is the same as in Fig. \ref{fig:S1FitnullE_AllBin}.}
	\label{fig:S1andS2Fit3kVcm_AllBin}
\end{figure}

\begin{figure*}[!tb]
	\begin{tabular}{c}
		\begin{minipage}{0.5\textwidth}
			\centering
			\includegraphics[width=1.0\hsize,clip]{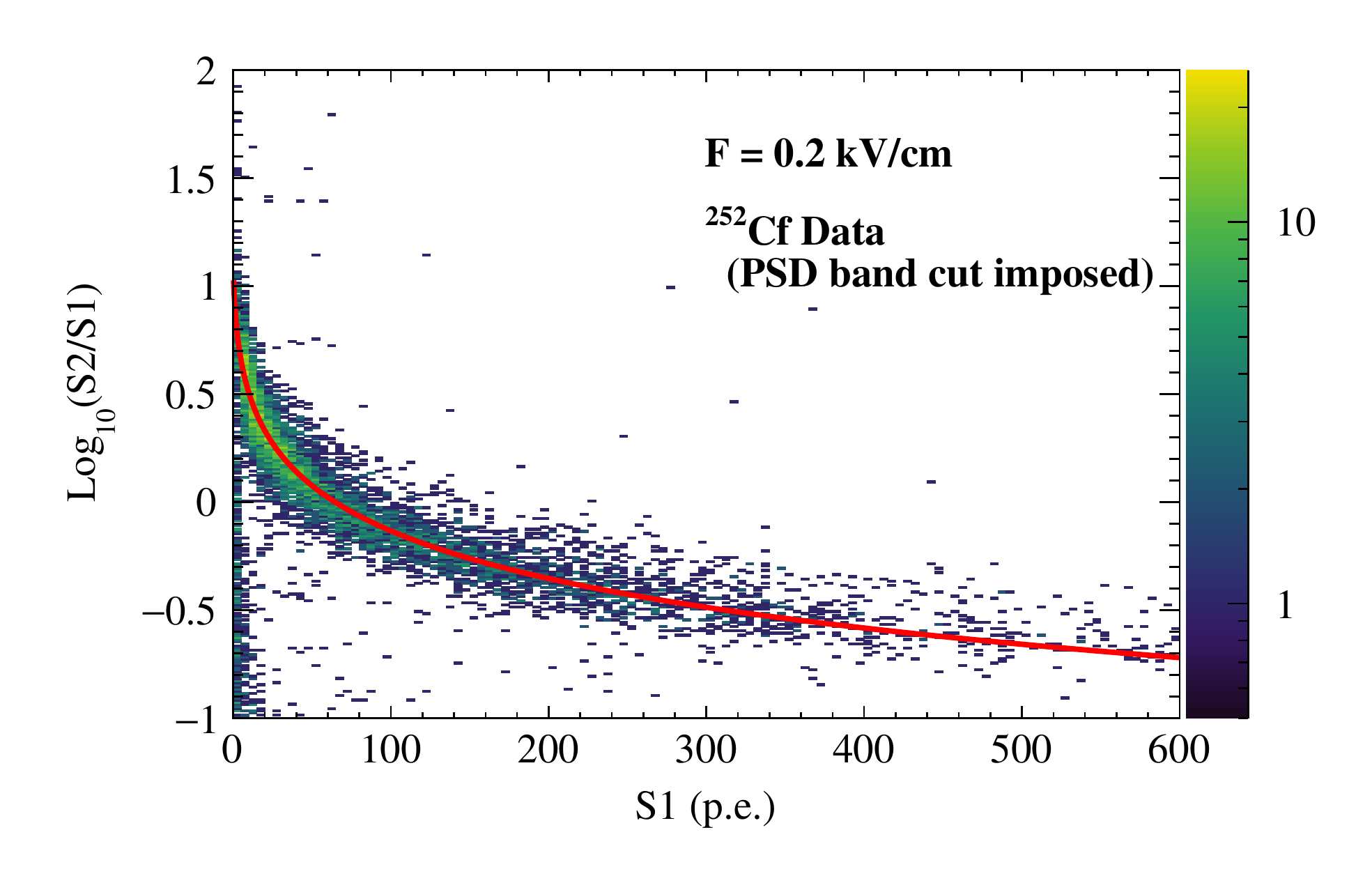}
		\end{minipage}
		\begin{minipage}{0.5\textwidth}
			\centering
			\includegraphics[width=1.0\hsize,clip]{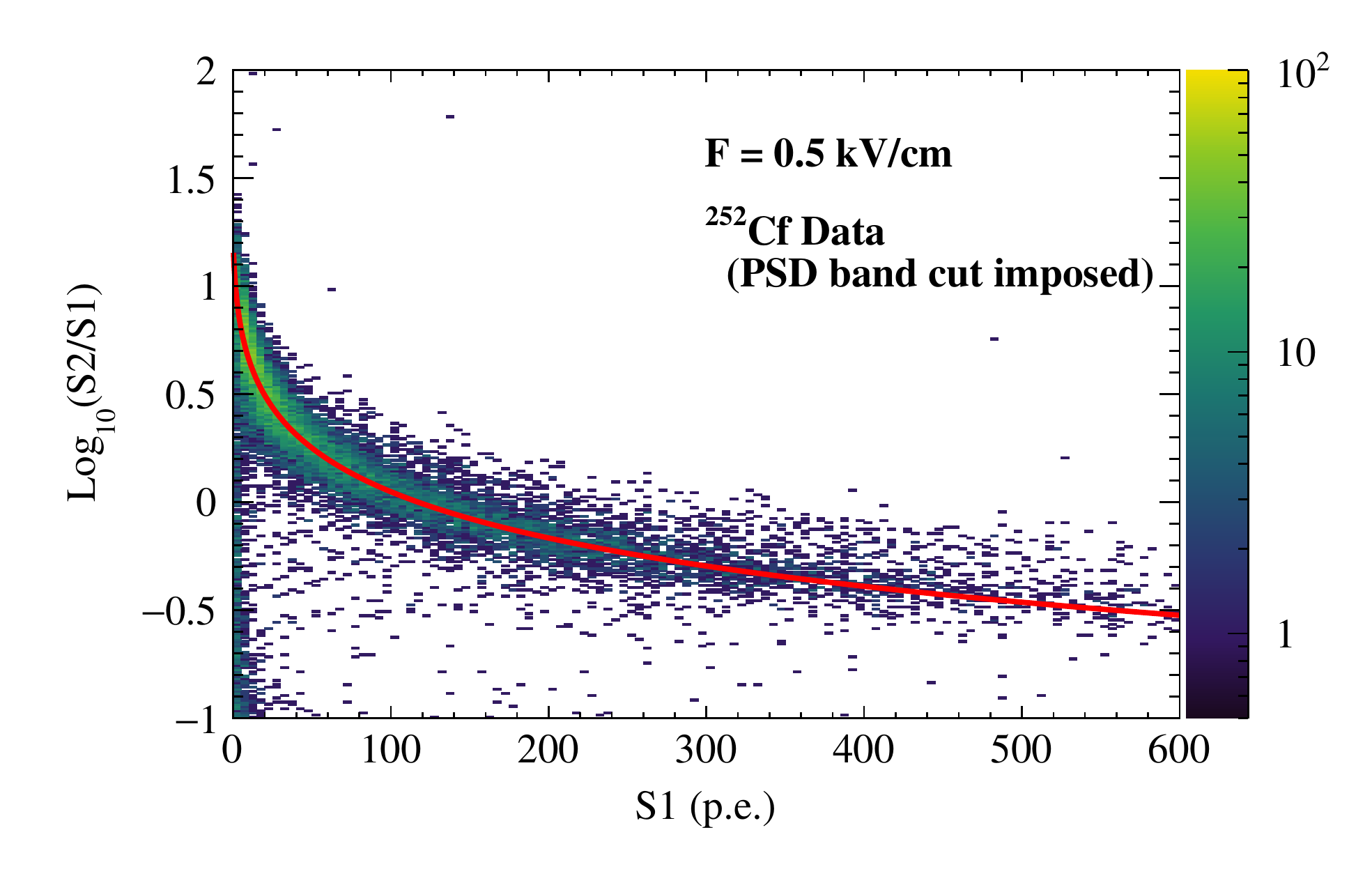}
		\end{minipage} \\
		\begin{minipage}{0.5\textwidth}
			\centering
			\includegraphics[width=1.0\hsize,clip]{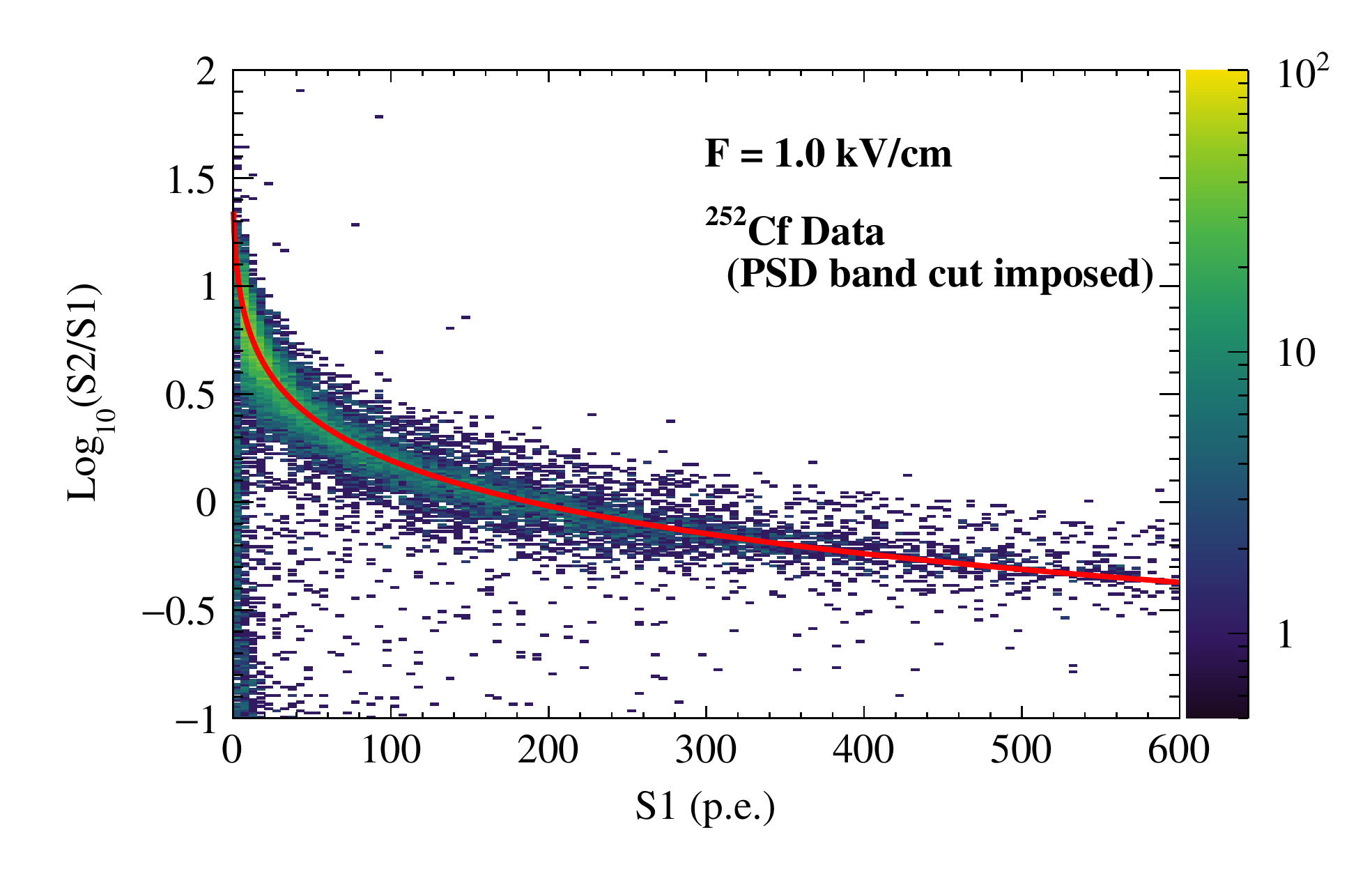}
		\end{minipage} 
		\begin{minipage}{0.5\textwidth}
			\centering
			\includegraphics[width=1.0\hsize,clip]{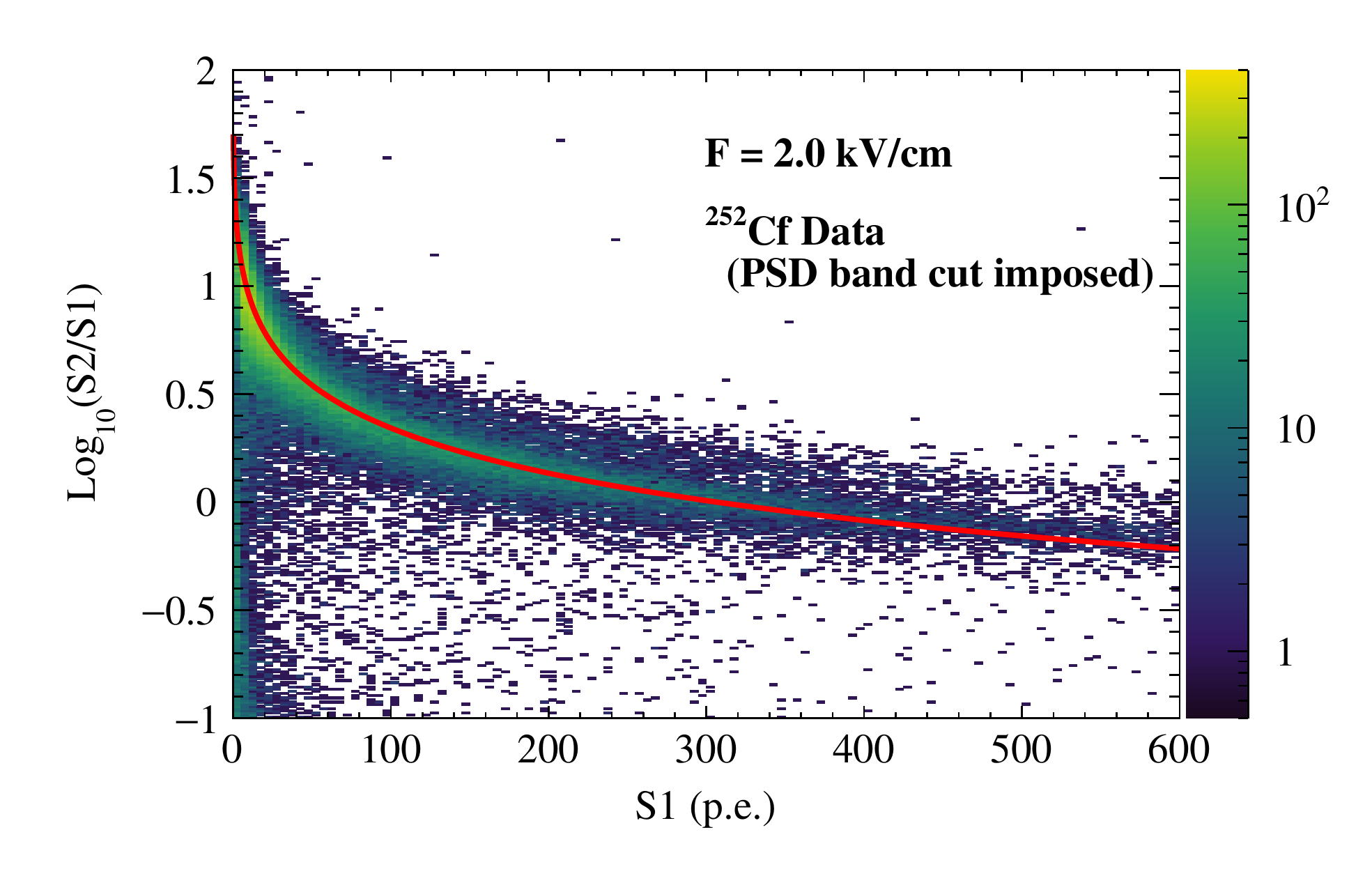}
		\end{minipage} \\
		\begin{minipage}{0.5\textwidth}
			\centering
			\includegraphics[width=1.0\hsize,clip]{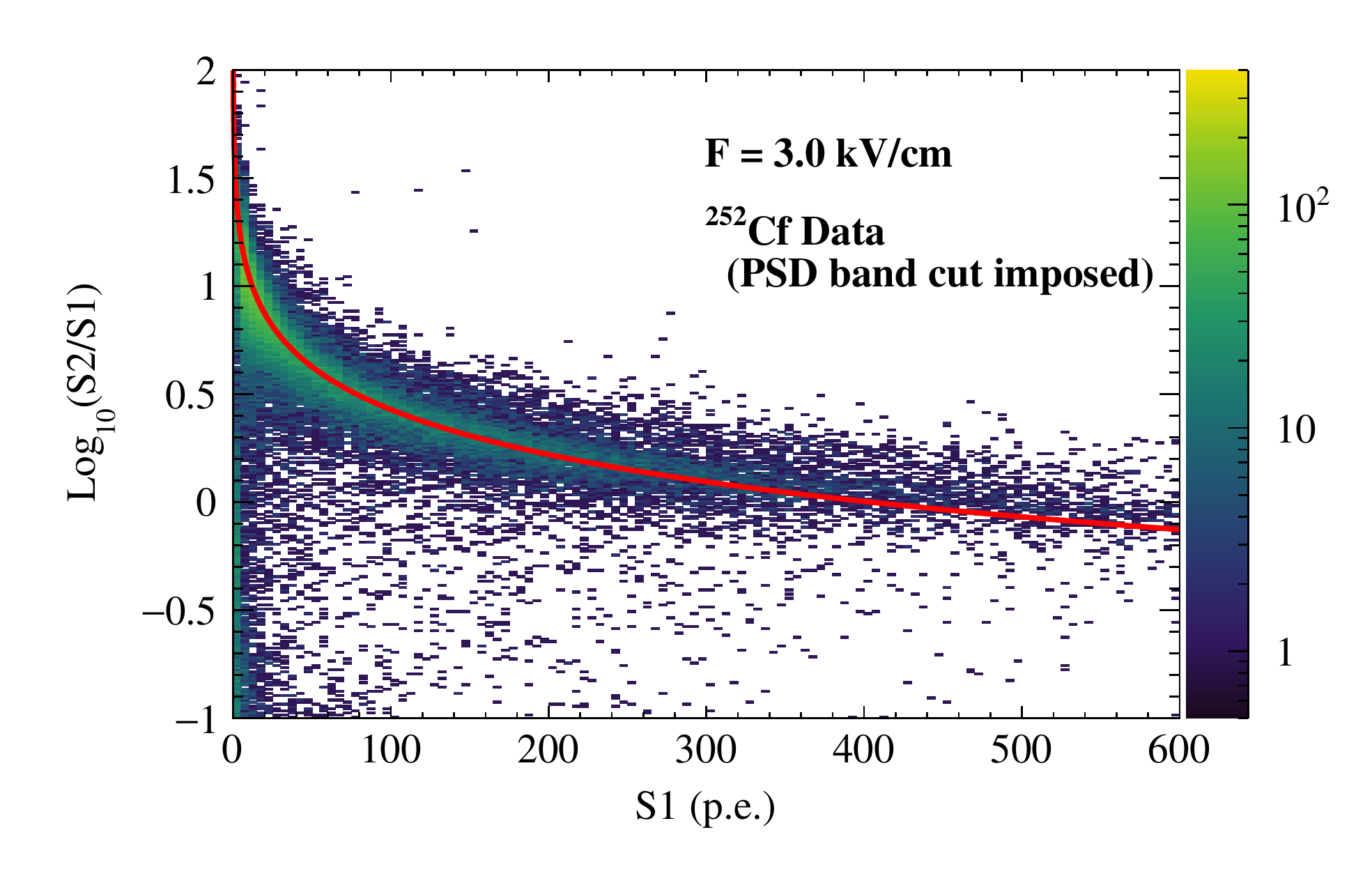}
		\end{minipage}
	\end{tabular}
	\caption{$\riCf$ data taken with the electric fields of $0.2, 0.5, 1.0, 2.0,$ and $3.0\ {\rm kV/cm}$ (from left to right and top to bottom) in $\s2s1$ versus the ${\rm S1}$ plane, overlaid with the prediction from the NR model and the best fit parameters (solid line).}
	\label{fig:S1vss2s1}
\end{figure*}

The parameters in the NR model are measured by fitting the obtained S1 and S2 spectra of each TOF bin ($4\ {\rm ns}$ interval) with the spectra derived from the MC simulation and the NR model described in Sec. \ref{sec:NRModel}.
The fit is simultaneously performed in the TOF range of $43$-$111\ {\rm ns}$ (total 17 TOF bins).
The MC spectra of both S1 and S2 are convolved with Gaussian resolution functions.
Figure \ref{fig:S1FitnullE_79_83ns} shows an example of the S1 spectrum and the fitted MC spectrum for a TOF bin of $79$-$83\ {\rm ns}$ at a null field.
Figure \ref{fig:S1FitnullE_AllBin} shows the spectrum for the entire TOF range of interest with the 17 MC spectra for each TOF bin.
Figure \ref{fig:S1andS2Fit3kVcm_79_83ns} shows an example of the S1 and S2 spectra and the fitted MC spectra for a TOF bin of $79$-$83\ {\rm ns}$ under the electric field of $3\ {\rm kV/cm}$.
The spectra for the entire TOF range of interest with the respective 17 MC spectra are shown in Fig. \ref{fig:S1andS2Fit3kVcm_AllBin}.
We should note that since the spectra for the entire TOF range of interest have a smooth spectrum shape, as shown in Figs. \ref{fig:S1FitnullE_AllBin} and \ref{fig:S1andS2Fit3kVcm_AllBin}, it is difficult to uniquely resolve the degeneracy between the free parameters by the inclusive shape.
However, the backscatter edge of each TOF bin makes it possible to access each parameter.
This is because the edges characterize a light and charge yield dependency on the NR energy. 
We also note that the fit range of the S2 spectra is constrained to below $500\ {\rm p.e.}$ as the discrepancy between data and the MC simulation is observed above $500\ {\rm p.e.}$
This discrepancy is presumably due to the multiple scattered events that survive the event selections mentioned above.

As a demonstration, Fig. \ref{fig:S1vss2s1} shows the $\riCf$ data for all the five values of an electric field with an overlay of the prediction (shown by solid line) in $\s2s1$ versus the ${\rm S1}$ plane.
Reasonable agreements of the mean value of the $\s2s1$ distributions are achieved at all the five electric fields.

\section{\label{sec:Result}Result}
\begin{table}[t]
	\caption{Results from the simultaneous fit of $\riCf$ data with the MC simulation and the NR model described in Sec. \ref{sec:NRModel}, together with their statistical uncertainties.}
	\begin{tabular*}{1.0\columnwidth}{@{\extracolsep{\fill}}ll} \hline
		Parameter 					& Value \\ \hline
		$k_B\ [{\rm g/(MeV \cdot cm^2)}]$ 	& $(3.12 \pm 0.05) \times 10^{-4}$ \\ 
		$\alpha_0 \ {\rm (fixed)}$ 			& $1.0$ \\
		$D_\alpha\ {\rm [(V/cm)^{-1}]}$ 		& $(8.9 \pm 0.5)\times 10^{-4}$ \\
		$\gamma\ {\rm [(V/cm)^{\delta}]}$ 	& $1.15 \pm 0.02$ \\
		$\delta$ 						& $(5.76 \pm 0.03)\times 10^{-1}$ \\ \hline
	\end{tabular*}
	\label{tab:fitResult}
\end{table}

\begin{figure}[tb]
	\centering
	\includegraphics[width=1.0\columnwidth]{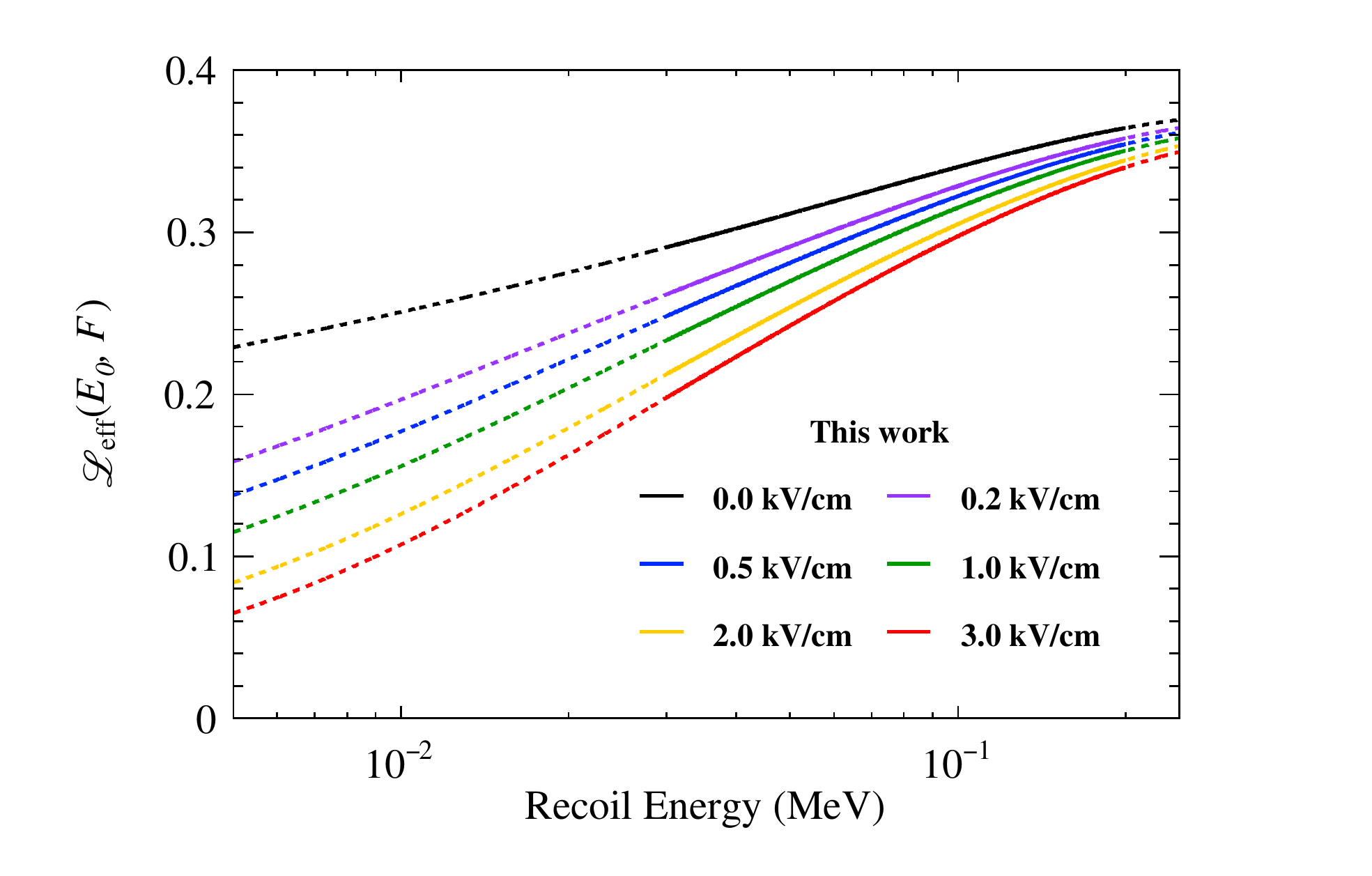}
	\caption{Scintillation efficiencies $\Leff$ as a function of the NR energy measured in this work.
	The colored solid lines represent the results from this work, and the corresponding dashed lines are extrapolations.}
	\label{fig:Leff}
\end{figure}

\begin{table}[t]
	\caption{Sets of fit parameters corresponding to four systematic uncertainty sources (energy calibration of the detector (E-calib.), the distance from TPC to $\riCf$ source (TOF-arm), TOF time calibration estimated by direct $\gamma$-ray events (TOF $t_{0}$), and the value of the $g_2/g_1$ ratio ($g_2/g_1$)).
	The units of each parameter are the same as in Table \ref{tab:fitResult}.}
	\begin{tabular*}{1.0\columnwidth}{@{\extracolsep{\fill}}llllll} \hline
		\multicolumn{2}{l}{\multirow{2}{*}{Source}}	& $k_B$			& $D_\alpha$ 		& \multirow{2}{*}{$\gamma$}	& $\delta$ \\
		&								& $[\times 10^{-4}]$	& $[\times 10^{-4}]$	&						& $[\times 10^{-1}]$ \\ \hline
		\multirow{2}{*}{E-calib.}& $+0.3\ {\rm p.e./keV_{ee}}$		& $3.71$			& $9.2$			& $1.15$					& $5.75$ \\
						& $-0.3\ {\rm p.e./keV_{ee}}$		& $2.54$			& $8.9$			& $1.16$					& $5.76$ \\ \hline 
		\multirow{2}{*}{TOF-arm}	& $+1\ {\rm cm}$	& $3.35$			& $8.2$			& $1.16$					& $5.77$ \\
						& $-1\ {\rm cm}$	& $2.90$			& $9.5$			& $1.15$					& $5.75$ \\ \hline
		\multirow{2}{*}{TOF $t_{0}$}	& $+1\ {\rm ns}$& $3.54$			& $8.9$			& $1.15$					& $5.75$ \\
									& $-1\ {\rm ns}$	& $2.71$			& $8.8$			& $1.15$					& $5.77$ \\ \hline
		\multirow{2}{*}{$g_2/g_1$}	& $+20\%$	& N/A			& $12$			& $0.93$					& $5.78$ \\
							& $-20\%$		& N/A			& $5.4$			& $1.62$					& $5.86$ \\ \hline
	\end{tabular*}
	\label{tab:fitUncertainty}
\end{table}

\begin{figure}[tb]
	\centering
	\includegraphics[width=1.0\columnwidth]{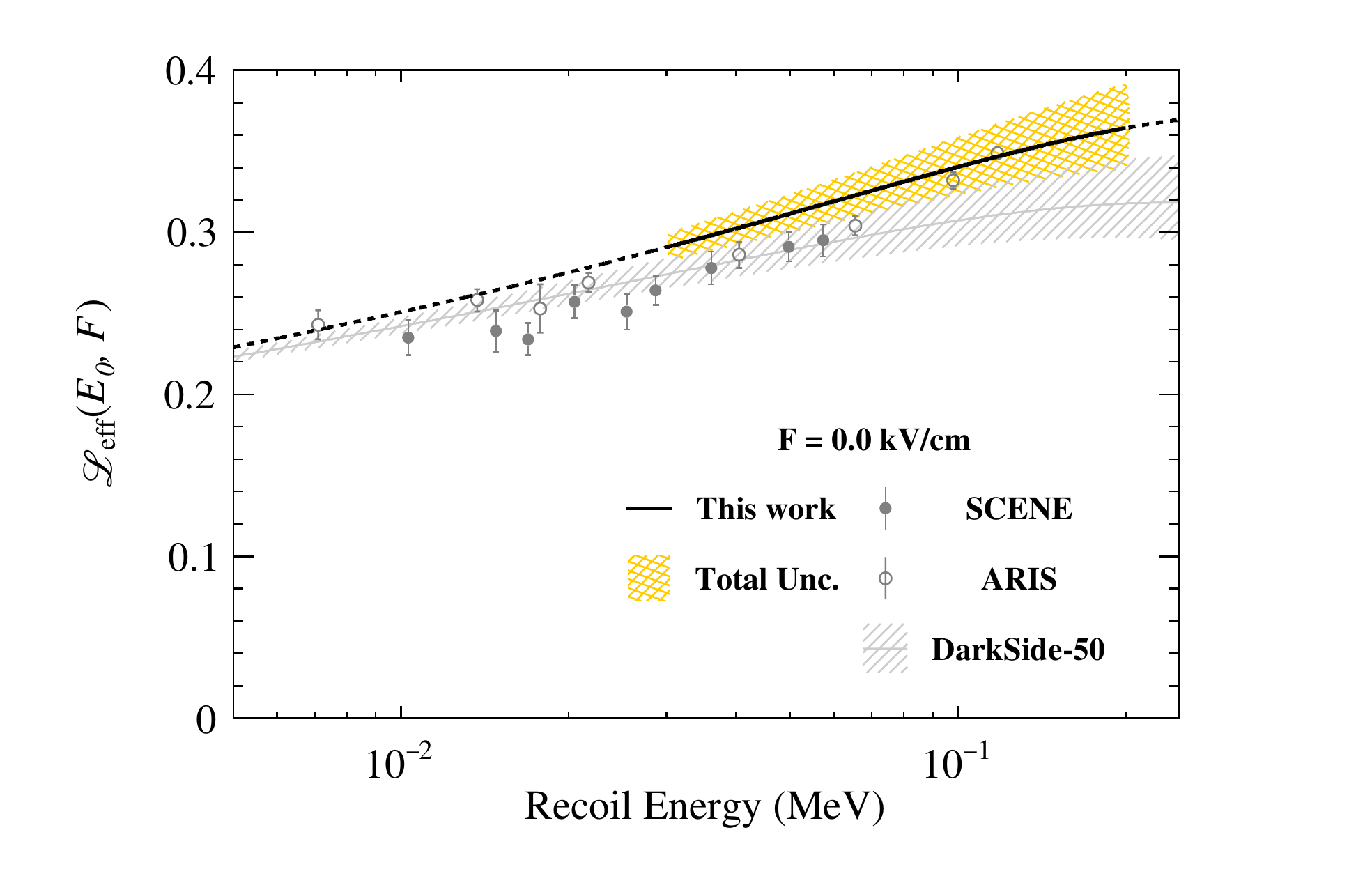}
	\includegraphics[width=1.0\columnwidth]{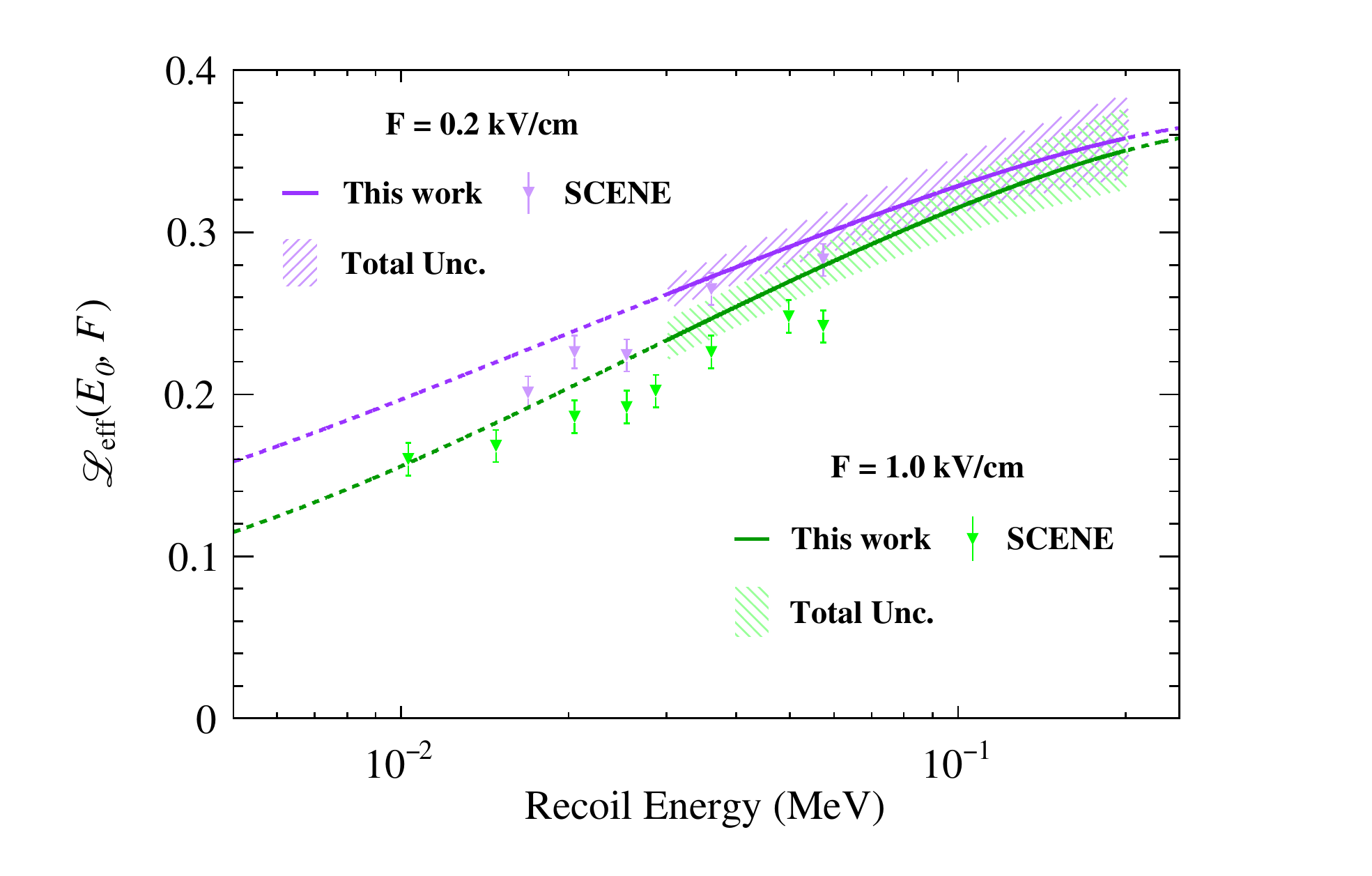}
	\caption{Top: Comparison with previous $\Leff$ measurements (SCENE \cite{cao2015measurement}, ARIS \cite{agnes2018measurement}, and DarkSide-50 \cite{agnes2017simulation}) at a null field.
	The black solid line is the result from this work (same as in Fig. \ref{fig:Leff}), and the orange band represents the total uncertainty on $\Leff$.
	Bottom: The comparison under electric fields of $0.2$ and $1.0\ {\rm kV/cm}$.
	The solid colored lines are the results from this work (same as in Fig. \ref{fig:Leff}), and 
	the corresponding bands represent total uncertainty on the $\Leff$, including the uncertainty from the $g_2/g_1$ ratio.}
	\label{fig:Leff_Sys}
\end{figure}

A set of the best fit parameters is summarized in Table \ref{tab:fitResult} and the resulting $\Leff$ spectrum at each electric field is shown in Fig. \ref{fig:Leff}.
We constrain the S1 fit range between $30\ {\rm keV}$ and $200\ {\rm keV}$ in order to have sufficient PSD power to extract pure NR events and also to ensure enough statistics for stabilizing the fitting procedure.
From the functional modeling, however, the energy range can be extrapolated to both lower and higher energy regions as represented with dashed lines in Fig. \ref{fig:Leff}.

In this measurement, there are four systematic uncertainties: energy calibration of the detector, the distance from TPC to $\riCf$ source, the absolute TOF measurement, and the $g_2/g_1$ ratio, which are considered to be uncorrelated each other.
We evaluate the impacts on the fitting parameters by shifting up/down within their uncertainties as shown in Table \ref{tab:fitUncertainty}.
In principle, all the data sets, i.e. all the electric fields data, are affected by these systematic sources in the same way, so we vary each uncertainty for all the data sets in common and reperform all the fitting.
It should be noted that uncertainty on energy calibration, mainly due to time dependence on the PMT gain and absolute light yield, is partially independent on different data sets, though we treat it as fully correlated to assign conservative error on this measurement.
The value $k_B$ is determined by S1 only with null field data; thus, it is not affected by $g_2/g_1$ uncertainty at all.
Statistical uncertainty throughout the measurement is about $10\%$-$20\%$ of the systematic uncertainty.
In addition, although we are not aware of any theoretical description of the empirical field dependency of $\alpha$, the model of Eq.(\ref{eq:alpha_def}) seems valid with our data samples and parametrizations.

Since the scintillation response in LAr for ER in the range $41.5$-$511\ {\rm keV}$ at a null field is constant within $1.6\%$ \cite{agnes2018measurement}, our result can be subjected to the  comparison with other $\Leff$ measurements using other reference sources (such as $^{83m}{\rm Kr}$ \cite{cao2015measurement} or $^{241}{\rm Am}$ \cite{agnes2018measurement}), a different experimental setup, and analysis method.
Figure \ref{fig:Leff_Sys} shows the comparison of $\Leff$ from this work to the previous measurements by other groups \cite{cao2015measurement, agnes2018measurement, agnes2017simulation} for without an electric field (top) and with electric fields (bottom) cases.
The colored bands in Fig. \ref{fig:Leff_Sys} represent the total uncertainties, evaluated by adding each deviation of $\Leff$ due to the systematic shift in Table \ref{tab:fitUncertainty} in quadrature.
Although this work shows systematically higher $\Leff$ than the other measurements, they are still consistent within their uncertainties.

The products from this measurement (parameter list, function form, and uncertainty band) are publicly available online \cite{kylablink}.

\section{\label{sec:Concl}Conclusion}
The scintillation efficiency of LAr for NR ranging from $30$ to $200\ {\rm keV}$ relative to $511\ {\rm keV}$ ER is systematically measured under electric fields from $0$ to $3\ {\rm kV/cm}$ using a small size double-phase TPC and a $\riCf$ radioactive source.
In this measurement, observed S1 and S2 spectra are simultaneously fit with the simulated energy deposits by taking into account correlations between the light and charge yields.
The parametrization model we employ in this paper is based on existing models (the Mei model and TIB model) described by the function of the recoil energy and electric field.
As a result, the scintillation efficiency $\Leff$ is successfully modeled up to $3\ {\rm kV/cm}$ within systematic uncertainty.
The model allows us to fully predict the scintillation yield at any recoil energy and any electric field between $0$ to $3\ {\rm kV/cm}$.

In the field of WIMP dark matter search experiments, the scintillation efficiency and the charge yield are essential parameters to convert from observed S1 and S2 signals to the recoil energy by WIMP-Ar scattering.
Thus, comprehensive parametrization of LAr property reported in this work makes use of interpretation between the experimental data and physics process under various electric fields, and also can contribute to better understand systematic uncertainty for a low energy signal region in the search.
Furthermore, not only a WIMP search, these results would be also useful for other physics experiments \cite{Ackermann2013the, Acciarri2016long, akimov2018coherent}, where understandings of the LAr property are necessary and play important role for obtaining physics outcomes.

\section*{\label{sec:Acknow}Acknowledgments}
The authors would like to thank Dr. Alan E. Robinson for helpful comment on the Geant4 simulation and its nuclear data library files.
This work is a part of the outcome of research performed under the Waseda University Research Institute for Science and Engineering (Project No. 2016A-507), supported by JSPS Grant-in-Aid for Scientific Research on Innovative Areas (Grants No. 15H01038/No. 17H05204), Grant-in-Aid for Scientific Research(B) (Grants No. 18H01234),  and Grant-in-Aid for JSPS Research Fellow (Grants No. 18J13018).
The authors acknowledge the support by Institute for Advanced Theoretical and Experimental Physics, Waseda University.

\nocite{*}

\bibliography{LAr}

\end{document}